\documentclass[12pt]{article}
\usepackage{amsmath,amssymb}
\usepackage{ifpdf}
\makeatletter

    \@addtoreset{equation}{section}
\makeatother
\ifpdf
  \usepackage[pdftex]{graphicx}
\else
  \usepackage[dvipdfm]{graphicx}
\fi

\newcommand{\ba}{\begin{eqnarray}}
\newcommand{\ea}{\end{eqnarray}}
\newcommand{\nn}{\nonumber}

\newcommand{\cO}{{\mathcal{O}}}
\newcommand{\cF}{{\mathcal{F}}}

\newcommand{\be}{\begin{equation}}
\newcommand{\ee}{\end{equation}}

\newcommand{\vev}[1]{ \left\langle {#1} \right\rangle }

\def\a{\alpha}
\def\b{\beta}

\def\l{\lambda}

\def\D{\Delta}

\def\L{\Lambda}

\allowdisplaybreaks[1]

\begin{document}

\vskip 12mm

\begin{center}
{\Large \bf Virasoro irregular conformal block and  \\
beta deformed random matrix model}
\vskip 1cm
{\large  Sang Kwan Choi, Chaiho Rim  and Hong Zhang}
\vskip 5mm
{\it Department of Physics and Center for Quantum Spacetime (CQUeST)}\\
{\it Sogang University, Seoul 121-742, Korea}
\end{center}

\vskip 10mm

\begin{abstract}
Virasoro irregular conformal block is presented 
as the expectation value of Jack-polynomials 
of the beta-deformed Penner-type matrix model
and  is compared with the  inner product of 
Gaiotto states with arbitrary rank.
It is confirmed that there are non-trivial modifications
of the Gaiotto states 
due to the normalization of the states.  
The relation between the two is
explicitly checked for rank 2 irregular conformal block. 

\end{abstract}

\vskip 12mm

\setcounter{footnote}{0}

\section{Introduction} 

Virasoro irregular module appears in connection with 
the N=2 super-Yang Mills theory \cite{G_2009}.
The irregular module so called Gaiotto state or Whittaker state \cite{Whittaker}
is the simultaneous eigenstate of the positive Virasoro generators.
The irregular module is constructed 
as the superposition of one primary state 
and its  descendents \cite {G_2009, MMM_2009}.

On the other hand, the irregular module is also constructed 
as the colliding limit of primary operators as shown in \cite{GT_2012}.
The colliding limit is the fusion of primary vertex operators 
with the addition of Heisenberg-coherent modes. 
As a result, the state becomes the simultaneous eigenstate 
of positive Virasoro operators, {\it i.e.}\/  the irregular module.

Will the two different approaches produce the same result? 
In this paper we like to answer this question. 
We will confine ourselves to the case with the Gaiotto sate $|I_n \rangle $ 
of rank $ n \ge1$, simultaneous eigenstate of  $L_n, L_{n+1}, \cdots, L_{2n}$.
In section 2, Gaiotto state of rank $n$ constructed in  \cite{KMST_2013}
 is summarized and its inner product  is investigated. 
The inner product is important since 
it contains all the information of descendents 
in the Gaiotto state. 
In section 3, a differently looking form of the inner product is provided
using  the colliding limit of the regular conformal correlation.
The result is given in terms of
the beta-deformed Penner-type matrix model.
Since the  random matrix model is the result of the fusion of primary operators,
the partition function should produce 
the colliding limit of the conformal block,
which we call  the (two-point) irregular conformal block (ICB).
A simple and clear way to obtain ICB is presented 
with the help of the loop equation
and ICB is compared with the inner product of Gaiotto states.
We pinpoint the non-trivial modification
from the Gaiotto state in \cite{KMST_2013}. 
The summary and discussion are given in section 4 
and some detailed calculation is given in the appendix.

\section{Virasoro irregular module and its inner product} 

The irregular state is explicitly constructed for rank 1 
in \cite{G_2009, MMM_2009} 
and  for  rank $n$ 
in \cite{KMST_2013}. 
We will use the convention  $| \widetilde {G_{2n}} \rangle$
for Gaiotto state with rank $n$ following  \cite{KMST_2013}
(another form is also found in \cite{BMT_2011}),
whereas we reserve $|I_n \rangle $ for the state obtained 
from the colliding limit given  in  \cite{GT_2012}.
\be
| \widetilde {G_{2n}} \rangle =
\sum_{\ell,Y, \ell_p} 
\L^{\ell/n}
\left\{ 
\prod_{i=1}^{n-1}  a_i ^{\ell_{2n-i} } b_i^{\ell_i}  
\right\} 
m^{\ell_n} 
Q_{\Delta} ^{-1} \Big(
1^{\ell_1}2^{\ell_2}
\cdots 
 (2n-1) ^{\ell_{2n-1}} (2n) ^{\ell_{2n}}; Y \Big)
L_{-Y} |\Delta \rangle \, ,
\label{G_2n}
\ee
where $L_{-Y}=L_Y^+ $ represents the product of lowering operators
and  $L_Y = L_1^{\ell_1} L_2^{\ell_2} \cdots L_s^{\ell_s}$. 
$ |\Delta \rangle $ is the primary state with conformal dimension $\Delta$
and $Q_{\Delta} (Y;Y')$ is the shorthand notation of $ \langle \Delta | L_{Y'} L_{-Y}|\Delta \rangle$. The summation $\ell$ runs from 0 to $\infty$, 
$Y$ and $\ell_p$ maintaining $|Y|=\ell$ and $\sum p \ell_p =\ell$.

One can confirm that $|\widetilde {G_{2n}} \rangle$ is the simultaneous eigenstate; 
$ L_{k}| \widetilde {G_{2n}} \rangle = \Lambda^{k /n} a_{2n-s} $  for $n < k \le 2n$
and  $  L_{n }  | \widetilde {G_{2n}} \rangle =\Lambda m    | \widetilde {G_{2n}} \rangle$
from the expectation values for $W=1^{\ell_1}2^{\ell_2}
\cdots   (2n) ^{\ell_{2n}}$,
\begin{align}
 &\langle \Delta| L_W L_{2n-s}| \widetilde {G_{2n}} \rangle 
=\Lambda^{2n-s /n} a_s    
 \langle \Delta| L_W | \widetilde {G_{2n}} \rangle     
~~~~{\rm  for} ~0 \le s <n, 
\nn\\
 &\langle \Delta| L_W L_{n}| \widetilde {G_{2n}} \rangle 
=\Lambda m  
 \langle \Delta| L_W | \widetilde {G_{2n}} \rangle \, ,
\end{align} 
with $a_0 \equiv1$.  
Here, the eigenvalues are given in terms of $\Lambda$, $a_i$'s and $m$ only.
The other coefficients $b_i$'s are not fixed by the eigenvalues 
but  enter in the inner product since 
\be 
\langle \Delta| L_W | \widetilde {G_{2n}} \rangle =\Lambda^{\ell /n}   
\left\{ \prod_{i=1}^{n-1}  a_i ^{\ell_{2n-i} } b_i^{\ell_i}  \right\} m^{\ell_n} \,.
\label{L+}
\ee
Note that inner product contains all the information on the descendents.
Thus, one may assume that $b_i$'s are related with the contribution of 
descendents. 
To find out  further information of $b_i$'s,  we need to resort to other 
procedures.

\section{Irregular conformal block and colliding limit} 

The inner product can be evaluated using 
the idea of colliding limit of the multi-point regular conformal correlation
introduced in \cite{EM_2009,GT_2012,BMT_2011}.
We follow the procedure appeared in  \cite{NR_2012}.
Let us consider the conformal part of $n+2$ primary operator correlation
with $N$ screening operators. 
If one fuses $n+1$ operators at the origin with the colliding limit,
one ends up with the $\beta$-deformed Penner-type partition function 
\be
Z_{(0:n)}  (c_0;  \{c_k\}) = \int \prod_{i=1}^N d\lambda_i ~\Delta(\lambda)^ {2 \beta} ~
e^{ -\frac{\sqrt{\beta}}g \sum_i V (\lambda_i ; c_0, \{c_k\}) } \, ,
\label{penner-0n}
\ee
where $\Delta(\lambda)=\prod_{i<j} (\lambda_i-\lambda_j)$ 
is the Vandermonde determinant and 
$\b = -b^2$ (or $b=i \sqrt{\b }$) with the screening charge $b$.
The Penner-type potential is given as the sum of logarithmic and 
inverse power terms
\be
\frac{1}\hbar V_{(0:n)} (z; c_0, \{c_k\}) = -c_0 \log z+  \sum_{k=1}^{n} \frac{c_k}{k z^k}\,.
\label{potential-0-n}
\ee 
(One may identify $c_k=  \sum_{r=1}^n \a_r (z_r)^k $
where $\a_r$ is the Liouville charge of the primary operator at $z_r$. 
Since the colliding limit corresponds to $z_r \to 0$ and 
$\a_r \to \infty$ so that $c_k$ is ensured finite,
one  may consider the limit as the ideal multi-pole expansion.
In addition, we  use the notation 
$g=i \hbar/2$ so that $\sqrt{\b }/g = -2b/\hbar$.)

We remark by passing that  the integration range of the partition function 
is naturally given as $0$ to $\infty$. 
Before the colliding limit 
one usually chooses the integration range 
between the positions of the primary operators.
For example, one may choose the position of the 
primary operators as  ($0, z_1=z, z_2=1, \infty$) and 
chooses the integration ranges from $0$ to $z_1$ or  from $z_2$ to $\infty$. 
However, to have the proper colliding limit, one needs to choose the integration 
range from $z_1$ to $z_2$ and take the limit  $z_1 \to 0$ and $z_2 \to \infty$. 

Let us introduce the primary state  $|\Delta\rangle$ 
with conformal dimension $\Delta = c_0  (Q-c_0)$
in the presence of the background charge  $Q$.
Then  $ \langle \Delta|$ is the primary state with  
the conformal dimension $\Delta = c_\infty  (Q-c_\infty)$
where $ c_\infty $ is fixed by the neutrality condition 
$c_0 + c_\infty +  bN =Q$. 
The colliding limit introduces the irregular state $ | I_n \rangle $ 
and the partition function is identified with the inner product 
$Z_{(0:n)}  (c_0;  \{c_k\})= \langle \Delta | I_n \rangle $.
This ensures that the irregular state $ | I_n \rangle $ is dependent on 
the set of coefficients $ \{c_1, \cdots c_n\}$.
In fact, it is demonstrated  in \cite{GT_2012} that the coefficient $c_k$ 
is  the coherent coordinate  of  Heisenberg mode $a_k$,
$a_k  | I_n \rangle = c_k  | I_n \rangle$.

Since $ | I_n \rangle $ is the simultaneous eigenstate of 
$L_n, L_{n+1}, \cdots, L_{2n}$ generators, 
their eigenvalues can be  parametrized as 
$\Lambda_k = (k+1) Q c_k - \sum_{p=0}^k c_p c_{k-p}$
with  $k=n, \cdots, 2n$. 
However, the eigenstate condition is not enough to fix $ | I_n \rangle$
as seen in \eqref{L+} and needs the information on 
the descendents in $| I_n \rangle $.
Note that the lower positive generators $L_k$ ($k=1, \cdots, n-1$)
obeying $ \left[L_k, L_n\right]=(k-n) L_{k+n} $. 
An easy way to realize this non-commutative properties 
is to represent $L_k$ as the differential form 
of the  coherent coordinates $c_k$'s. 
Putting $\mathcal{L}_k = \Lambda_k + v_k$, 
one has 
\be
v_k \equiv \sum_{\ell \in \mathbb{N}} \ell ~ c_{\ell+k} ~\frac{\partial}{\partial c_{\ell}} 
\label{v_k}
\ee
and the consistency condition
\be
[v_k,  v_{\ell} ] \langle \Delta | I_n \rangle 
= (\ell-k) ~v_{\ell+k} \langle \Delta | I_n \rangle \, .
\label{consistency}
\ee
It should be noted that the Gaiotto state
$ | \widetilde {G_{2n}} \rangle $ in \eqref{G_2n}
satisfies the consistence condition trivially since 
$v_k \langle \Delta  | \widetilde {G_{2n}} \rangle  =0$.

One can find the parameter dependence for the  rank 1 
simply by scaling the integration variable $\lambda_i \to c_1 \lambda_i $ 
to get $Z_{(0:1)}  (c_0; c_1) =c_1^{- b N  (  b N + 2 c_0-Q) } Z_{(0:1)}  (c_0; 1)$. 
However, for the rank higher than 1, one needs more complicated process. 
The easiest way to find the parameter dependence 
is to use the loop equation of the matrix model. 
The loop equation has the form \cite{NR_2012}
\be 
\sum_{k=0}^{n-1} \frac{v_k ( \log(Z_{(0:n)}))}{z^{2+k}} =
-\frac{ \xi(z) }{\hbar^2}\, ,
\label{loop}
\ee
where $v_0 $ conforms to the notation of \eqref{v_k}, $v_0 
\equiv \sum_{\ell \in \mathbb{N}} \ell \, c_{\ell } \, \frac{\partial}{\partial c_{\ell}} $
and $ \xi(z) =4 W(z)^2 -4 W(z)   V'(z) + 2 \hbar Q  W'(z) -\hbar^2 W(z,z) $.
Here $W(z)$ is the resolvent $W(z) = \hbar b/2 \langle 1/(z-\lambda_i)\rangle$, 
and $W(z,z)$ is the connected two-point resolvent 
$W(z,z)= -b^2 \langle \sum_{i,j} 1/(z-\lambda_i)(z-\lambda_j)\rangle_c$.
The  prime stands for the differentiation. 

One may view that the loop equation provides
the energy momentum  expectation value $\varphi_2(z)$, 
which encodes the Seiberg-Witten curve \cite{SW1, SW2, r:DF,G_2009}.
Putting
$\varphi_2(z)= \sum_{n\le k \le 2n} \Lambda_k /z^{2+k} + \sum_{0\le k n-1} {\cal L}_k /z^{2+k}$,
one has the relation with the resolvent according to the loop equation:
$\varphi_2(z) = (2 W -V')^2 + \hbar Q (2 W -V')' - \hbar^2 W(z,z)$. 
Large $z$ expansion of the loop equation eventually reduces to the flow equation 
\be
v_k (\log Z_{(0:n)})   = d_k^{(0:n)} (\{c_k\})\, ,
\ee
where $d_k^{(0:n)}$ is the moment of  $\xi(z)$; 
$\oint dz z^{1 +k} \xi(z)  /(-\hbar^2 2 \pi i) $. 
The flow equation satisfies the consistency condition 
\eqref{consistency} automatically whose explicit solutions can be found 
in \cite{NR_2012,CR_2013}. 

\begin{figure}
\centering
\includegraphics[width=0.8\textwidth]{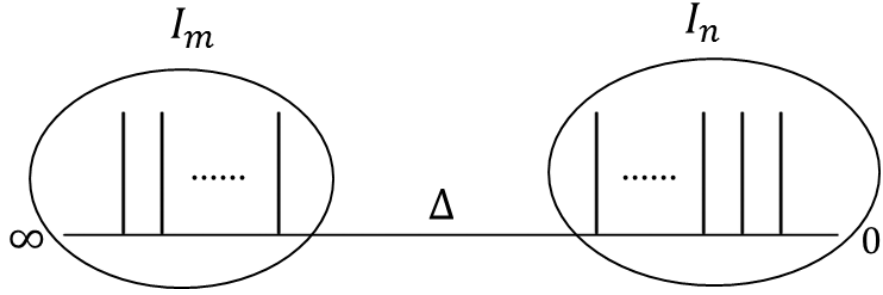}
\caption{Schematic diagram of $\vev{I_m |I_n}$ from the colliding limit}
\label{f:irregular}
\end{figure}

The idea can be extended to find the inner product $\langle I_m |I_n \rangle $ 
from the colliding limit of $(m+n +2)$-point correlation (see figure \ref{f:irregular}).
Fusing  $n+1$ primary operators at the origin and $m+1$ operators at infinity, 
one has the partition function  $ Z_{(m:n)} $ 
\begin{align}
& 
Z_{(m:n)}  (c_0; \{c_k\}; \{c_{-\ell}\}) = \int \prod_{i=1}^N d\lambda_i \Delta(\lambda)^ {2 \beta} 
e^{ -\frac{\sqrt{\beta}}g \sum_i  V_{(m:n)}  (\lambda_i; c_0, \{c_k\} , \{c_{-\ell}\}) } \, ,
\nn\\
&
 \frac 1 \hbar V_{(m:n)}   (z; c_0, \{c_k\} , \{c_{-\ell}\})
= -c_0 \log z + \sum_{k=1 }^{n}   \left( \frac {{ c_k} } {k z^{k}} \right)   
 + \sum_{\ell=1}^{m}   \left( \frac { c_{-\ell}~ z^{\ell}}   {\ell}  \right )  \,.
\label{Z_mn}
\end{align}  
The partition function is related with 
the inner product $\langle I_m | I_n \rangle$. 
However, there is a  subtlety, so called $U(1)$ contribution. 
This factor comes from the limiting procedure:
It is noted that  as $z_a \to \infty$
and $z_b \to 0$  one has the finite contribution 
$\prod_{a,b}  (1 -z_b/z_a)^{-{2 \alpha_a \alpha_b}} \to e^{\zeta_{(m:n)}} $, 
where $\zeta_{(m:n)} =\sum_k^{{\rm min}(m,n)} 2 c_k c_{-k}/k$.
Therefore, one has the inner product  of the form
$ \langle I_m| I_n \rangle = e^{\zeta_{(m:n)}}   Z_{(m:n)}  (c_0; \{c_k\}; \{c_{-\ell}\}) $.

The inner product between the two irregular modules inherits the property of the 
conformal block of the regular multi-correlation. 
Considering the colliding limit, one may define the irregular conformal block
 ${\cal F}_\Delta^{(m:n)}  $ 
as the inner product of the irregular modules
with appropriate
normalization: 
${\cal F}_\Delta^{(m:n)}   ={ \langle I_m | I_n \rangle }/({ \langle I_m| \Delta \rangle ~ \langle \Delta | I_n \rangle})$
whose conformal dimension  is given as 
$\Delta= (c_0 +N_0 ) (Q- c_0-N_0 ) = (c_\infty+ N_\infty) (Q-c_\infty -N_\infty) $
\cite{CR_2013}.

In this spirit, one may naturally 
define ICB using the $\b$ deformed Penner-type matrix model as
the following:
\be
{\cal F}_\Delta^{(m:n)}  (\{c_{-\ell }\} : \{ c_k \}) = 
\frac{e^{\zeta_{(m:n)}}   Z_{(m:n)}  (c_0; \{c_k\}; \{c_{-\ell}\})} 
{Z_{(0:n)}(c_0;  \{ c_k \})  Z_{(0:m)} ( c_\infty; \{ c_{-\ell }\}) }\,,
\label{ICB-m}
\ee 
where  $Z_{(0:n)}(c_0;  \{ c_k \})  $ and $Z_{(0:m)} ( c_\infty; \{ c_{-\ell }\}) $ 
provide the proper normalization for the irregular conformal block.
Here we use the change of variable $\lambda_i \to 1/\lambda_i$ to 
express $ \langle I_m| \Delta \rangle $ as $ Z_{(0:m)} ( c_\infty; \{c_{-\ell }\})$.

To evaluate ICB we note that the potential $V_{(m:n)}  $
contains the information of the irregular module at the origin and at infinity 
at the same time. Therefore, each module can be derived if one views the same potential 
on a different footing. 
The information of the irregular module at the origin is obtained 
if one regards  the potential  $V_0 = V _{(0:n)} (\{\lambda_i \} ; c_0, \{c_k\})  $ 
as the reference one and $\Delta V _0$ as its perturbation:
\be 
 \frac 1 \hbar V_0 
=\sum_{I=1} ^{N_0} \Big( 
 -c_0 \log \lambda_I +  \sum_{k=1}^{n} \frac{c_k} {k } \lambda_I^{-k} \Big)
\,;~~~ 
 \frac 1 \hbar \Delta V _0  =\sum_{I=1} ^{N_0} \Big( 
  \sum_{\ell=1}^{n} \frac { c_{-\ell}}{\ell}  { \lambda_I^{\ell}}  \Big) \,.
\ee
That is, $V_0 $ is the potential for the partition function $Z_{(0:n)} $ with 
$N_0 (\le N)$  number of screening operators. 
At infinity one has  the reference potential 
$ \sum_{J=1} ^{N_\infty} \Big( 
 -c_0 \log \lambda_J  + \sum_{\ell=1}^{n} { c_{-\ell}}{ \lambda_i^{\ell}} /\ell  \Big)$ 
and its perturbation $  \sum_{J=1} ^{N_\infty }  
\Big( \sum_{k=1}^{n} {c_k}  \lambda_J^{-k}/k \Big) $.
We introduce the number  $N_\infty $ of screening operators  at infinity 
so that  $N_\infty  +N_0 =N$.
One may rewrite the potential in a familiar form if one changes the variable 
$\lambda_J \to 1/\mu_J$ to get the equivalent potential
\be 
  \frac 1 \hbar V_\infty = \sum_{J=1} ^{N_\infty} \Big( 
 -c_\infty \log \mu_J  + \sum_{\ell=1}^{m} \frac { c_{-\ell}} {\ell}  \mu_{\!J} ^{-\ell}  \Big)
\,; ~~~~
 \frac 1 \hbar \Delta V _\infty =  \sum_{J=1} ^{N_\infty }  
\Big( \sum_{k=1}^{n} \frac {c_k} {k}  \mu_{\! J}^k  \Big) \,.
\ee
In this way the perturbative potential and the cross terms
 in the Vandermonde determinant provide ICB : 
\be
{\cal F}_\Delta^{(m:n)}  (\{c_{-\ell }\} : \{ c_k \}) = 
e^{\zeta_{(m:n)}} 
\Big \langle 
\prod_{I, J}  (1 - \lambda_I \mu_J)^{2\beta}  
 e^{ -\frac{\sqrt{\beta}}g (\Delta V_0 (\lambda_I) + \Delta V_\infty (\mu_J) ) }
\Big \rangle\, ,
\label{ICB}
\ee
where the bracket denotes the expectation value using the reference partition function: 
\begin{align}
&\langle {\mathcal O}(\lambda_I) \rangle 
\equiv \langle {\mathcal O}  \rangle_+ 
= \Big ( Z_{(0:n)}(c_0;  \{ c_k \}) \Big)^{-1} \int \prod_{I=1}^{N_0} d\lambda_I \Delta(\lambda)^ {2 \beta} 
 {\mathcal O}(\lambda_I) 
e^{ -\frac{\sqrt{\beta}}g \sum_i  V_0  (\lambda_I)  }  \, ,
\\
&\langle {\mathcal O}(\mu_J) \rangle 
\equiv\langle {\mathcal O}  \rangle_-
= \Big ( Z_{(0:m)}(c_\infty;  \{ c_\ell \}) \Big)^{-1} \int \prod_{J=1}^{N_\infty} d\mu_J \Delta(\mu)^ {2 \beta} 
 {\mathcal O}(\mu_J) 
e^{ -\frac{\sqrt{\beta}}g \sum_i  V_\infty  (\mu_J)  } \nn \,,
\end{align}
which can be regarded as the generalization of Selberg integral \cite{0710.3981, Forrester}.
One may put ICB in \eqref{ICB} compactly 
in terms of Jack polynomial \cite{Kadell, Macdonald}. 
Putting $p_k =\sum_I \lambda_I^k $ and $p'_k =\sum_J \mu_{\! J}^k$,
one has the identity 
\begin{align}
& \prod_{I, J}  (1 - \lambda_I \mu_J)^{2\beta}  
 e^{ -\frac{\sqrt{\beta}}g (\Delta V_0 (\lambda_I) + \Delta V_\infty (\mu_J) ) }
\nn\\
&~~~~~~~~~~=
\exp\big\{-\b\sum_{k=1}^{\infty} \frac{1} {k } p_k (p'_k-\tilde c_{-k}) \big\} \times
\exp\big\{-\b\sum_{k=1}^{\infty} \frac{1} {k } p'_k (p_k-\tilde c_{k}) \big\}\, ,
\end{align}
where $\tilde c_{\pm k }=i2c_{\pm k}/\sqrt{\b}  = -2 c_{\pm k}/ b$, (and $\tilde c_{k}=0$ for $k>n$ and $\tilde c_{-k}=0$ for $k>m$).
Using the Cauchy-Stanley identity \cite{1012.3137, Stanley}
$ e^{\b \sum_{k \ge 1} \frac 1k p_k p'_k } 
=\sum_R j_R ^{(\b)}  (p) j_R^{(\b)} (p') $,
one has  ICB as  
\be
 {\cal F}_\Delta^{(m:n)}  = 
e^{\eta_{(m:n)}}   \sum_{Y,W}
\Big \langle 
j_Y ^{(\b)}  (p_k) j_W^{(\b)} (-p_k+\tilde c_{k})
\Big \rangle_+
\Big \langle 
j_Y ^{(\b)} (-p'_k+\tilde c_{-k}) j_W^{(\b)} (p'_k)
\Big \rangle_-\,\,.
\label{ICB-Jack}
\ee

The explicit form of the general ICB  is not available yet. 
Here we check a few non-trivial terms  
using the resolvent in the loop equation 
of the reference partition function. 
Each term can be obtained from the large $z$ expansion of the resolvent $W(z)$.
The details of calculation are given in the appendix. 
ICB is given in power of $\eta_0 \equiv c_1 c_{-1}$,
which is compatible with the Young diagram expansion.
For the rank 1, up to order  $ \cO (\eta_0^2 )$ one has 
\be 
{\cal F}_\Delta ^{(1:1)}=1+\eta_0  \frac{2\bar{c_0} \bar{c_\infty}}{\D} 
+\eta_0^2\frac{ {4\bar{c_0}^2\bar{c_\infty}^2 c}/{\D} 
+4\D+2+12(\bar{c_0}^2+\bar{c_\infty}^2)
+32\bar{c_0}^2\bar{c_\infty}^2}{c+2 c \D+2 \D(8 \D-5)}\, ,
\ee
where $\bar{c_0}=Q-c_0$, $\bar{c_\infty}=Q-c_\infty$,
$c=1+6Q^2$.  
Comparing this with 
the Gaiotto inner product  up to $\cO (\L  \L')^2$ 
(using \eqref{G_2n}  with $\langle \widetilde{G}_2 | $ using the primed notation)
\be
\langle \widetilde{G_2}|\widetilde{G_2} \rangle=
1+\L \L' \frac{m m'}{2\D}
+ (\L \L')^2
\frac{  {m^2 m'^2 c}/{4\D} +4\D+2-3 (m^2+m'^2)+2 m^2 m'^2 }{c+2 c \D+2 \D(8 \D-5)}\, ,
\ee
we find  $\L^2= - c_1^2$ and $m \L=2 c_1 \bar{c_0}$, consistent with the eigenvalues of $L_2$ and $L_1$.

Non-trivial check is given for the rank 2.
Matrix model provides ${\cal F}_\Delta ^{(1:2)}$ up to $\cO\left(\eta_0^2 \right)$ 
\begin{align}
& {\cal F}_\Delta ^{(1:2)}=
1+\eta_0 \frac{\bar{b}_1 \bar{c_\infty}}{\D} 
\nn\\
& ~~~ + \eta_0^2   \frac {c \, \bar{c_\infty}^2 \, \bar b_2/\Delta 
+c_2(2+12 \bar{c_\infty}^2+4\D)(1-(Q+2\bar{c_0})/c_1^2)
+(3+8 \bar{c_\infty}^2)\,
\bar b_2}  {c+2 c \D+2 \D(8 \D-5)}\, ,
\label{f12}
\end{align}
where $c_1 \bar{b}_1=(d_1^{(0:2)} +2 \bar{c_0}c_1)$,
$c_1^2 \, \bar b_2= (d_1^{(0:2)} +2 \bar{c_0}c_1)^2 + 
c_2 \, \partial\,  d_1^{(0:2)}/ \partial c_1 $.
The explicit form of $d_1^{(0:2)}$ is found in \eqref{d02}. 
On the other hand, 
one has the Gaiotto inner product up to $\cO (\L' \sqrt{ \L}  )^2 $
\be
\langle \widetilde{G_2}|\widetilde{G_4} \rangle=1+  \L' \sqrt{ \L}  
\frac{ m'  b_1}{2 \D}
+ (\L' \sqrt{ \L}  )^2 \frac{{c  m'^2 b_1^2}/(4\D) +(2-3m'^2+4\D) m+(-3+2 m'^2) b_1^2}
{c+2 c \D+2 \D(8 \D-5)}\,.
\label{g2g4}
\ee
Comparing the two 
we obtain the parameter relations 
$\L'^2= - c_{-1}^2$, $m'  \L'=2 c_{-1} \bar{c_\infty}$ and $ \L m = -c_1^2-2 c_2 (\bar{c_0}-Q/2) $,
the eigenvalues of $L_1$, $L_2$.
However $\bar{b}_2$ is not $\bar{b}_1^2$ in \eqref{f12} which is different from \eqref{g2g4}.
Therefore, $(\sqrt{\L} \, b_1)^\ell$ cannot be considered as a simple constant but 
should be of the form
$ (c_1 \, \bar b_1)^\ell =\frac1{Z_{(0:2)}}(\L_1+v_1)^\ell Z_{(0:2)} $.

One can check this relation holds for  $\cF_\D^{(2:2)}$
if the Gaiotto inner product 
\be
\begin{split}
\langle \widetilde{G_4}|\widetilde{G_4} \rangle
=1&+(\L'  \L)^{1/2} \frac{ b_1 b'_1}{2 \D} 
+\frac{\L' \L}{c+2 c \D+2 \D(8 \D-5)} \biggl[ \frac{ c \, b_1^2 {b'}_1^2 }{4\D} \\
&+2(b_1^2 {b'}_1^2+m  m')
-3( m {b'}_{1}^2+3m' b_1^2)
+4 \D m m'
\biggr]+(\L' \L)^{3/2} 
\end{split}
\ee
is compared with the matrix result given in \eqref{F22}.
Additional identification of  $b_1'$ with $\bar b_{-1}$ 
appears as it should be, 
where 
$ (c_{-1} \, \bar b_{-1})^\ell =\frac1{Z_{(0:2)}}(\L_{-1}+v_{-1})^\ell Z_{(0:2)} $.

\section{Summary and discussion} 

We found the Virasoro  irregular conformal block
using the beta deformed Penner type matrix model
and present the result in terms of the expectation values of 
the Jack polynomial \eqref{ICB-Jack}. 
We check ICB explicitly for a few ranks
and compare with the inner product of Gaiotto state 
proposed by \cite{KMST_2013}.
There is a non-trivial modification 
between the two results 
due to the difference of the normalization
as is suggested in \cite{KMST_2013}. 

Referring to the explicit check given for the rank 1 and 2,
we can clearly see that the Gaiotto state needs to be modified
to represent the colliding limit of the conformal correlation. 
Note that the expectation value  $\langle \Delta| L_k^{\ell_k} | \widetilde {G_{2n}} \rangle $ 
is  $(\Lambda^{k/n} b_k)^{\ell_k}$ according to \eqref{L+}.
On the other hand, $| I_n\rangle$ has the $L_k^{\ell_k}$ expectation 
value $ \Big( ( \Lambda_k + v_k)^{\ell_k}  Z_{(0:n)} \Big)/ Z_{(0:n)}$.
Therefore, one may conclude that the Gaiotto state should be 
modified by using the coefficient relation 
$(\Lambda^{k/n} b_k)^{\ell_k} \to \Big( ( \Lambda_k + v_k)^{\ell_k}  Z_{(0:n)} \Big)/ Z_{(0:n)} $.
Considering $\langle \Delta| L_1^{\ell_1} L_2^{\ell_2} | I_n \rangle 
=(\L_2+v_2)^{\ell_2}(\L_1+v_1)^{\ell_1} Z_{(0:n)} $, one has
\be
\langle \Delta| L_W | I_n \rangle
=  \Lambda^{\ell/n}  m^{\ell_n} 
\left\{\prod_{i=1}^{n-1}  a_i ^{\ell_{2n-i} }\right\} 
\left\{ \left(\L_{n-1}+v_{n-1}\right)^{\ell_{n-1}}  
\cdots \left(\L_{1}+v_{1}\right)^{\ell_{1}}  
{Z_{(0:n)}}\right\}
\ee
with proper ordering.
The case of rank 1 is trivial since there is no $b_k$'s.

In the paper we consider mainly the two-point ICB.
One may extend the result to $N$-point ICB 
$\langle \prod_{A=1}^N  I_{m_A}(z_A)  \rangle$
by generalizing the potential in \eqref{Z_mn}:
\be
 \frac 1 \hbar V_{(\{m_A\})}   (z; \{c_0^{(A)}\}, \{c_k^{(A)}\})
= \sum_{A=1}^N 
\left\{  -c_0^{(A)} \log (z-z_A) + \sum_{k=1 }^{n_A} 
  \left( \frac {{ c_k^{(A)}} } {k (z-z_A)^{k}} \right)   
\right\}\,.
\ee
ICB will be given with the appropriate normalization at each point,
{\it i.e.}, by treating the potential as the sum of  the reference potential 
and perturbation at each point. 

Noting the Penner-type matrix model provides ICB, 
one may wonder if there exists another systematic way of obtaining
the irregular conformal block of arbitrary rank 
from regular conformal block, 
as seen  in  the rank 1 case \cite{MMM_2009}
or for $SU(N)$ in \cite{MRZ_2014}
by decoupling a certain large mass limit. 
However, such a decoupling limit is not achieved yet
for rank greater than $1$.
It will be interesting to find the limit using 
the relation of the Selberg integral with the Jack polynomials
to have \eqref{ICB-Jack}. 

In addition, one may have the colliding limit for W-algebraic symmetry as done 
in \cite{KMST_2013} using $SU(N)$ Toda theories. 
The corresponding matrix model is straight-forward generalization 
of the Virasoro symmetric case 
 for $SU(N)$. Making use of \cite{Zhang:2011au}, we have
\begin{align}
& 
Z_{(m:n)}^{SU(N)}  = \int \prod_{a=1}^{N-1}\prod_{i=1}^{N_a} d\lambda_i^{(a)} \Delta(\lambda^{(a)} )^ {2 \beta} \prod_{a=1}^{N-2}\Delta(\lambda^{(a)},\lambda^{(a+1)} )^ {- \beta}
e^{ -\frac{\sqrt{\beta}}g \sum_{a=1}^{N-1}\sum_{i=1}^{N_a}  V_{(m:n)}^{(a)}  (\lambda_i) } \,,
\nn\\
&
 \frac 1 \hbar V_{(m:n)}^{(a)}   (z)
= -c_0^{(a)} \log z + \sum_{k=1 }^{n}   \left( \frac {{ c_k^{(a)}} } {k z^{k}} \right)   
 + \sum_{\ell=1}^{m}   \left( \frac { c_{-\ell}^{(a)}~ z^{\ell}}   {\ell}  \right )\,,
\label{V-toda}
\end{align}  
with  $c_k^{(a)}=  \sum_{r=1}^n (\a_r, e_a) (z_r)^k $, and $c_{-\ell}^{(a)}=  \sum_{r=1}^m (\tilde \a_r, e_a) (\tilde z_r)^{-\ell} $. Here $e_a$ are the simple roots of $SU(N)$. This leads to the $SU(N)$ ICB:
\begin{equation}
 {\cal F}_\Delta^{(m:n)} = 
e^{\zeta_{(m:n)}} 
\sum_{\vec{Y}}
\left\langle
\prod_{a=1}^{N}
j^{(\beta)}_{Y_a} (p_k^{(a-1)} -p_k^{(a)} + \tilde c_{k}^{(a)})
\right\rangle_{\! \! +}
\left\langle
\prod_{a=1}^{N}
j^{(\beta)}_{Y_a} (p_k'^{(a)} -p_k'^{(a-1)} +\tilde c_{-k}^{(a)})
\right\rangle_{\! \! -}
\end{equation}
with $ \tilde c_{-k}^{(a)}=2\sum_{s=1}^{a-1} c_{-k}^{(s)}/b $,  $ \tilde c_{k}^{(N-a)}=-2\sum_{s=1}^{a-1} c_{k}^{(N-s)}/b $, $p_k^{(0)}=p_k^{(N)}=p_k'^{(0)}=p_k'^{(N)}\equiv 0$, and $\langle {\mathcal O}  \rangle_{\pm}$ are generalizations of $A_{N-1}$ Selberg integral,
expectation values of the matrix model with the reference potential,
part of \eqref{V-toda}. $U(1)$ factor $\zeta_{(m:n)}$ is also summed over $SU(N)$ index  $a$.

\subsection*{Acknowledgements}
This work was supported by the National Research Foundation of Korea(NRF) grant funded by the Korea government(MSIP) (NRF-2014R1A2A2A01004951).

\appendix

\section{Matrix result for rank $1$}
Let us calculate ICB $\cF_\Delta^{(1:1)}$ defined in \eqref{ICB}
\be
{\cal F}_\Delta^{(1:1)} = 
e^{2c_1 c_{-1}} 
\Big \langle 
\prod_{I, J}  (1 - \lambda_I \mu_J)^{2\beta}  
 e^{ -\frac{\sqrt{\beta}}g (\Delta V_0 (\lambda_I) + \Delta V_\infty (\mu_J) ) }
\Big \rangle\, ,
\ee
where $\D V_0=\sum_{I=1}^{N_0} c_{-1} \l_I$ and $\D V_\infty=\sum_{J=1}^{N_1} c_{1} \mu_J$.
Rescaling the integration variables $\l_I \to c_1 \l_I$ and $\mu_J \to c_{-1} \mu_J$,
 ICB is expanded in powers of $\eta_0 \equiv c_1 c_{-1}$
\be
\begin{split}
e^{-2 \eta_0} \cF_\Delta^{(1:1)}
&=1+\eta_0 \left[ 2b \Bigl( \vev{\mu_J}+\vev{\l_I} \Bigr)
+2b^2 \vev{ \l_I} \vev{ \mu_J}\right] 
+\eta_0^2 \Bigl[ 2 b^2 \Bigl( \vev{\l_I \l_J} \\
&+2 \vev{\l_I}
\vev{ \mu_J}+\vev{ \mu_I  \mu_J} \Bigr)
+4 b^3 \Bigl(\vev{\l_I \l_J} \vev{\mu_k}+\vev{\l_k}\vev{\mu_I \mu_J} \Bigr) \\
&+b^2 \vev{\l_I^2} \vev{\mu_J^2}+2 b^4 \vev{\l_I \l_J} \vev{\mu_k \mu_\ell} \Bigr]
+\cO(\eta_0^3) \,.
\end{split}
\ee
Here we omit a summation symbol for a notational simplicity.
The expectation values can be obtained
using the loop equation which for $Z_{(0:n)}(\l_I)$ (or $Z_{(0:m)} (\mu_J)$)
\be
V'(z)^2+f(z)-\hbar Q V''(z)=(2 W(z)-V'(z))^2+\hbar Q (2 W(z)-V'(z))'-\hbar^2 W(z,z)\, ,
\ee
where $f(z)=2\hbar b \vev{\sum_i(-V'(z)+V'(\l_i))/(z-\l_i)}= 
\sum_k v_k (-\hbar^2 \log Z)/z^{2+k}  $ as in \eqref{loop}
where $Z$ can be either  $Z_{(0:n)}$ or $Z_{(0:m)}$. 
Expanding the resolvent in powers of $1/z$,
we have at the order of $z^{-3}$
\be
 b \vev{\l_I}=-\frac{b N_0}{ b N_0-\bar{c_0}}=\frac{\bar{c_0}-(a+\frac{Q}2)}{a+\frac{ Q}2}\, ,
\ee
where $\bar{c_0}$ denotes $Q-c_0$ and $a=Q/2-c_0-b N_0$ so that $\Delta=(Q/2)^2-a^2$.

At the order of $z^{-4}$, one has
\be
0=(b \vev{\l_I})^2+2 b \vev{\l_I}+2  b \left( b N_0-\bar{c_0} -\frac Q 2 \right) \vev{\l_I^2}
+ b^2 \vev{\l_I \l_J}_{\mathrm{c}} \,.
\label{(1:1)loop4}
\ee
To find $\vev{\l_I \l_J}_{\mathrm{c}}$,
one needs another identity for the resolvents \cite{NR_2012,E_2004}
\be
-\frac{\hbar^2}{4}W(z,z,z)+(2W(z)+V'(z))W(z,z)
+ \frac{W''(z)}{2}-U(z,z)+\frac{\hbar Q}4 W'(z,z)=0 \, ,
\label{wzzz}
\ee
where $W(z,z,z)$ is a $3$-point resolvent and
\be
U(z,z)=\beta \left\langle \sum_I \frac{-V'(z)+V'(\l_I)}{z-\l_I} \sum_J \frac1{z-\l_J} \right \rangle_{\mathrm{c}} \,.
\ee
Additional identities
$\beta \left\langle \sum_I \frac1{\l_I} \sum_J \frac1{z-\l_J} \right \rangle_{\mathrm{c}} = -\frac{\partial W(z)}{\partial c_1}
$ and
$\beta \left\langle -\sum_I V'(\l_I) \sum_J \frac1{z-\l_J} \right \rangle_{\mathrm{c}} = W'(z)
$ allow us to have $
U(z,z)=-\frac{W'(z)}{z}+\frac{c_1}{z^2}\frac{\partial W(z)}{\partial c_1} \,.$
Then at the order of $z^{-5}$ in \eqref{wzzz}, we have
\be
0=-b^2 ( b N_0-\bar{c_0}) \vev{ \l_I \l_J}_{c}+\frac{b}{2} \vev{\l_I^2} \,.
\label{(1:1)loop5}
\ee
From the two equations \eqref{(1:1)loop4} and \eqref{(1:1)loop5},
we find
\be
\begin{split}
& b \vev{\l_I^2}=-\frac{(a+\frac{Q}2-\bar{c_0})(a+\frac{Q}2+\bar{c_0})}
{(a+\frac{ Q}2)\Bigl(2(a+ Q)(a+\frac{ Q}2)+\frac{1}2\Bigr)} \,, \\
&b^2 \vev{\l_I \l_J}_{c}=\frac{(a+\frac{Q}2-\bar{c_0})(a+\frac{Q}2+\bar{c_0})}
{2(a+\frac{ Q}2)^2\Bigl(2(a+ Q)(a+\frac{ Q}2)+\frac{1}2\Bigr)} \,.
\end{split}
\ee

Similarly, we have the expectation values of $\mu_J$
by changing the parameters
$c_0 \to c_\infty$ and $N_0 \to N_\infty$,
which implies $a \to -a$ and $\bar{c_0} \to \bar{c_\infty}$ 
in the above result for $\l_I$.
Finally, we obtain
\be
\begin{split}
{\cal F}_\Delta ^{(1:1)}=1+\eta_0  \frac{2\bar{c_0} \bar{c_\infty}}{\D} 
&+\frac{\eta_0^2}{c+2 c \D+2 \D(8 \D-5)}
\biggl[\frac{4\bar{c_0}^2\bar{c_\infty}^2 c}{\D} \\
&+4\D+2+12(\bar{c_0}^2+\bar{c_\infty}^2)
+32\bar{c_0}^2\bar{c_\infty}^2
 \biggr]+\cO\left(\eta_0^3 \right)\, ,
\end{split}
\label{F^(1:1)}
\ee
where $c=1+6Q^2$.

\section{Matrix result for rank $2$}
Explicit form of  $Z_{(0:2)}$ is given in \cite{NR_2012, CR_2013}
where the expansion  in $\eta_1\equiv c_2/c_1^2$  is used 
to solve the loop equation.
The resolvent has two cuts and its solution is given 
in terms of filling fractions $N_{(1)}$ and $N_{(2)}$ 
with $N_0=N_{(1)}+N_{(2)}$.
One finds  
\be
\frac{d_1^{(0:2)}}{c_1}=-2 b N_{(2)} c_0
+\big[ -b N_0(b N_0 +2  c_0 -Q  )+b N_{(2)}(3b N_{(2)}+4 c_0-3Q) \big]\eta_1+\cO(\eta_1^2)\,.
\label{d02}
\ee

ICB $\cF_\D^{(1:2)}$ has the same form 
as $\cF_\D^{(1:1)}$ in \eqref{F^(1:1)}
except for one additional term $c_2 b \vev{\mu_J^2}/c_1^2$.
The loop equation shows at the order of $1/z^{3}$ and $1/z^{4}$
\begin{align}
-\frac{d_1^{(0:2)}}{c_1}&=2 b  N_0+2 b (b  N_0-\bar{c_0}) \vev{\l_I} \label{(1:2)loop3}\, , \\
0&=2  b  N_0 \eta_1+( b \vev{\l_I})^2+2  b \vev{\l_I}+2  b ( b  N_0-\bar{c_0}-\frac{Q}2) \vev{\l_I^2}
+ b^2 \vev{\l_I \l_J}_{c} \,.
\label{(1:2)loop4}
\end{align}
Using additional relation $
\beta \left\langle -\sum_I \l_I V'(\l_I) \sum_J \frac1{z-\l_J} \right \rangle_{c} = 
W(z)+z W'(z)$, 
we have $
U(z,z)=-\frac{W(z)}{z^2}-\frac{W'(z)}{z}+\frac{c_2}{z^3}\frac{\partial W(z)}{\partial c_1}$.
Then \eqref{wzzz} becomes at the order of $z^{-5}$
\be
0=-b^2 ( b N_0-\bar{c_0}) \vev{ \l_I \l_J}_{c}+\frac{b}{2} \vev{\l_I^2}
-\eta_1 \frac{\partial}{\partial c_1} (c_1 \vev{\l_I}) \,.
\label{(1:2)loop5}
\ee
Then \eqref{(1:2)loop3}, \eqref{(1:2)loop4} and \eqref{(1:2)loop5} solve
the expectation values as
{\footnotesize
\be
\begin{split}
& b \vev{\l_I}=-\frac{a+\frac{Q}2-\bar{c_0}}{a+\frac{Q}2}+\frac{D_1}{2(a+\frac{Q}2)} \,, \\
& b \vev{\l_I^2}=\frac{-4\left(\left(a+\frac{Q}2 \right)^2-\bar{c_0}^2\right)
+D_1^2+4 \bar{c_0} D_1+\left( B_1
-2\left(a+\frac{Q}2-\bar{c_0}\right)\left(4\left(a+\frac{Q}2\right)^2+1\right)\right)\eta_1}
{2\left(a+\frac{Q}2\right)\left(4\left(a+Q\right)\left(a+\frac{Q}2\right)+1 \right)} \,, \\
& b^2 \vev{\l_I \l_J}_{c}=
-\frac{1}{a+\frac{Q}2}\left(\frac{b \vev{\l_I^2}}2
+\frac{2\left(a+\frac{Q}2-\bar{c_0} \right)-B_1}{4\left(a+\frac{Q}2\right)}\eta_1\right)\, ,
\end{split}
\label{rank2ev}
\ee}
where $D_1\equiv d_1^{(0:2)}/c_1$ and $B_1\equiv \partial d_1^{(0:2)}/\partial c_1$.
Therefore, we obtain ICB as
\be
\begin{split}
{\cal F}_\Delta ^{(1:2)}&=1+\eta_0 \frac{\bar{D}_1 \bar{c_\infty}}{\D} 
+\frac{\eta_0^2}{c+2 c \D+2 \D(8 \D-5)}
\biggl[\frac{c \, \bar{c_\infty}^2 (\bar{D}_1^2+\bar{B}_1 \eta_1)}
{\D} \\
&+(2+12 \bar{c_\infty}^2+4\D)(1-(Q+2\bar{c_0})\eta_1)+(3+8 \bar{c_\infty}^2)
(\bar{D}_1^2+\bar{B}_1 \eta_1) \biggr]
+\cO\left(\eta_0^3 \right)\, ,
\end{split}
\label{F^(1:2)}
\ee
where $\bar{D}_1 \equiv D_1+2\bar{c_0}$ and
$\bar{B}_1 \equiv B_1+2\bar{c_0}$.

$\cF_\D^{(2:2)}$ can be evaluated
with additional terms of $\eta_1 b \vev{\mu_J^2}+\eta_{-1}b \vev{\l_I^2}$
where $\eta_{-1}=c_{-2}/c_{-1}^2$.
The expectation values of $\mu_J$ are obtained from \eqref{rank2ev}
by changing the parameters as
$a \to -a$, $\bar{c_0} \to \bar{c_\infty}$, $D_1 \to D_{-1}$ and $B_1 \to B_{-1}$.
Finally, we have
\be
\begin{split}
&\cF_\D^{(2:2)}=1+\frac{\bar{D}_1\bar{D}_{-1}}{2\D} \eta_0 \\
&+\frac{\eta_0^2}{c+2 c \D+2 \D(8 \D-5)}
\biggl[\frac{c (\bar{D}_1^2+\bar{B}_1\eta_1)(\bar{D}_{-1}^2+\bar{B}_{-1} \eta_{-1})}
{4\D} \\
&+2((\bar{D}_1^2+\bar{B}_1\eta_1)(\bar{D}_{-1}^2+\bar{B}_{-1} \eta_{-1})
+ (1-(Q+2 \bar{c_0})\eta_1)(1+(2 c_\infty-Q) \eta_{-1})) \\
&+3((1-(Q+2 \bar{c_0})\eta_1)(\bar{D}_{-1}^2+\bar{B}_{-1} \eta_{-1})
+(1-(Q+2 \bar{c_\infty}) \eta_{-1}) (\bar{D}_1^2+\bar{B}_1\eta_1))\\
&+4\D (1-(Q+2 \bar{c_0})\eta_1)(1-(Q+2 \bar{c_\infty}) \eta_{-1}) \biggr]
+\cO(\eta_0^3)\, ,
\end{split}
\label{F22}
\ee
where $\bar{D}_{-1} \equiv D_{-1}-2\bar{c_\infty}$
and $\bar{B}_{-1} \equiv B_{-1}-2\bar{c_\infty}$.



\begin{thebibliography}{99}

\bibitem{G_2009}
D. Gaiotto, {\it Asymptotically free N=2 theories and irregular conformal blocks}
[arXiv:0908.0307].

\bibitem{Whittaker}
E. Felinska, Z. Jaskolski, and M. Kosztolowicz, {\it Whittaker Pairs for the Virasoro Algebra and
the Gaiotto - Bmt States, } J. Math. Phys. 53 (2012) 033504, arXiv:1112.4453 [math-ph].

\bibitem{MMM_2009}
A. Marshakov, A. Mironov and  A. Morozov, 
{\it On non-conformal limit of the AGT relations},
 Phys. Lett. {\bf B682} (2009) 125-129 
[arXiv:0909.2052].


\bibitem{GT_2012}
D. Gaiotto and J. Teschner, {\it Irregular singularities in Liouville theory}, JHEP {\bf 1212}  (2012) 050
[arXiv:1203.1052]. 


\bibitem{KMST_2013}
H. Kanno, K. Maruyoshi, S. Shiba, and M. Taki,
{\it W$_3$ irregular states and isolated N=2 superconformal field theories}, JHEP {\bf 1303}  (2013) 147
[arXiv:1301.0721].


\bibitem{BMT_2011}
Giulio Bonelli, Kazunobu Maruyoshi, Alessandro Tanzini,
{\it Wild Quiver Gauge Theories}, JHEP {\bf 1202}  (2012) 031
[arXiv:1112.1691].

\bibitem{EM_2009} 
T. Eguchi and K. Maruyoshi, 
{\it Penner Type Matrix Model and Seiberg-Witten Theory}, JHEP {\bf 1002}  (2010) 022
[arXiv:0911.4797].

\bibitem{NR_2012}
T. Nishinaka and C. Rim,
{\it Matrix models for irregular conformal blocks and Argyres-Douglas theories}, JHEP {\bf 1210}  (2012) 138
[arXiv:1207.4480].

\bibitem{SW1}
 N.~Seiberg and E.~Witten,
 {\it Monopole Condensation, And Confinement In N=2 Supersymmetric Yang-Mills
  Theory, }
  Nucl. Phys. B {\bf 426} (1994) 19 
  [arXiv:hep-th/9407087].

\bibitem{SW2}
  N.~Seiberg and E.~Witten,
  {\it Monopoles, duality and chiral symmetry breaking in N=2 supersymmetric
  QCD, }
  Nucl.\ Phys.\  B {\bf 431} (1994) 484 
  [arXiv:hep-th/9408099].

\bibitem{r:DF}
  R.~Dijkgraaf and C.~Vafa,
  {\it  Toda Theories, Matrix Models, Topological Strings, and N=2 Gauge Systems, }
  [ arXiv:0909.2453 [hep-th]].


\bibitem{CR_2013}
S.-K. Choi and C. Rim,
{\it Parametric dependence of irregular conformal block}, JHEP {\bf 1404}  (2014) 106
[arXiv:1312.5535].

\bibitem{E_2004} 
B. Eynard,
{\it Topological expansion for the 1-Hermitian matrix model correlation functions},
 JHEP {\bf 0411} (2004) 031 [hep-th/0407261].



\bibitem{0710.3981}
  P. J. Forrester and S. O. Warnaar,
  {\it  The importance of the Selberg integral, }Bull. Amer. Math. Soc. (N.S.) 45 (2008) 489-534,
  arXiv:0710.3981 [math.CA].

\bibitem{Forrester}
P. J. Forrester, {\it Log-gases and random matrices, }volume 34 of London Mathematical Society
Monographs Series. Princeton University Press, Princeton, NJ, 2010.

\bibitem{Kadell}
K.W.J.Kadell,
{\it An integral for the product of two Selberg-Jack symmetric functions,  }
Compositio Math. {\bf 87} (1993) 5-43;
{\it  The Selberg-Jack symmetric functions,  }
Adv.Math. {\bf 130} (1997) 33-102.

\bibitem{Macdonald}
I. G. Macdonald, {\it Symmetric functions and Hall polynomials,   }Oxford Univ. Press, 1995.

\bibitem{1012.3137}
A.Mironov, A.Morozov and Sh.Shakirov,
{\it A direct proof of AGT conjecture at $\beta=1$ , }
JHEP {\bf 1102} (2011) 067[ arXiv:1012.3137 [hep-th] ].

\bibitem{Stanley}
R. P. Stanley, {\it Some combinatorial properties of Jack symmetric functions, }
Adv. Math. {\bf 77}
(1989) 76-115.

\bibitem{MRZ_2014} 
Y.~Matsuo, C.~Rim and H.~Zhang,
{\it Construction of Gaiotto states with fundamental multiplets through Degenerate DAHA}, 
JHEP  {\bf 1409} (2014) 028
[arXiv:1405.3141 [hep-th] ]. 

\bibitem{Zhang:2011au} 
  H.~Zhang and Y.~Matsuo,
  {\it Selberg Integral and SU(N) AGT Conjecture, }
  JHEP {\bf 1112}  (2011) 106
  [arXiv:1110.5255 [hep-th] ]. 






\end{thebibliography}
\end{document}